\documentclass[pre,10pt,amsmath,amssymb]{revtex4}
\usepackage{graphicx}

\usepackage{amsmath,amsfonts,amssymb,bm}

\begin{document}

%\draft
\title{Multiplicative L\'evy noise in bistable systems}

\author
{Tomasz Srokowski}

\affiliation{
 Institute of Nuclear Physics, Polish Academy of Sciences, PL -- 31-342
Krak\'ow,
Poland }

%\date{\today}

\begin{abstract} 

Stochastic motion in a bistable, periodically modulated potential is discussed. 
The system is stimulated by a white noise increments of which have a symmetric 
stable L\'evy distribution. The noise is multiplicative: its intensity depends 
on the process variable like $|x|^{-\theta}$. The Stratonovich 
and It\^o interpretations of the stochastic integral are taken into account. The mean 
first passage time is calculated as a function of $\theta$ for different values of 
the stability index $\alpha$ and size of the barrier. Dependence of the output amplitude  
on the noise intensity reveals a pattern typical for the stochastic resonance. 
Properties of the resonance as a function of $\alpha$, $\theta$ and size of 
the barrier are discussed. Both height and position of the peak strongly depends 
on $\theta$ and on a specific interpretation of the stochastic integral. 

\end{abstract} 


\maketitle

\section{Introduction}

Ubiquity of the normal distribution in nature is a consequence of its stability. However, 
stable distributions are not restricted to the Gaussians: processes which involve 
discontinuous trajectories and algebraic tails $\sim|x|^{-1-\alpha}$, 
where $0<\alpha<2$ is a stability index, can be stable as well. Divergent moments 
of those distributions reflect a presence of long jumps (L\'evy flights). The normal distribution 
corresponds to $\alpha=2$. The L\'evy stable statistics with $\alpha<2$ is important 
when one considers realistic phenomena, where complexity, non-uniformity and long-range 
correlations play a role. For that reason, they are encountered and studied, 
for example, in biology \cite{wes}, sociology \cite{broc} 
and finance \cite{mant,san}. A characteristic feature of many problems in those fields 
is a non-homogeneous structure of the medium which results in an anomalous transport, as it is 
the case for porous, disordered materials \cite{bou} and folded polymers \cite{bro}. The non-homogeneity 
of the medium has significant consequences for a stochastic description of such systems in terms 
of the Langevin equation; the noise can no longer be additive and must include a dependence 
on the process variable. Despite that, multiplicative problems with L\'evy flights are rarely 
discussed in the literature. 

It is a well-known fact that the Langevin equation with the multiplicative white noise is not unique 
and requires an additional interpretation of the stochastic integral. Two of them are 
of particular interest: the It\^o interpretation (II), for which the integrand 
is evaluated before the noise action, and the Stratonovich one (SI) 
which takes into account the process value 
both before and after the stochastic stimulation. SI is distinguished since it constitutes 
a limit of the coloured noise in a form of the fractional Ornstein-Uhlenbeck process \cite{sron}. 
On the other hand, II applies for discrete systems and systems with a large inertia 
\cite{kup,sron}. II leads to a fractional Fokker-Planck equation with a variable diffusion 
coefficient \cite{sche} and it differs from that for SI only by a spurious drift if $\alpha=2$. 
For non-Gaussian processes a relation between both interpretations is more 
complicated and predictions qualitatively different: for SI, in contrast to II, 
slope of the distribution tail is modified by the multiplicative factor  \cite{sro1,sro2}. 
In particular -- in the case of the linear driving force -- that factor may sufficiently 
weaken the noise to make the variance of the output signal finite. Slopes corresponding to 
the finite inertia lie between values for II and SI \cite{sron} which emphasises importance 
of taking into account both interpretations. 

One also can achieve the finite variance by introducing a nonlinear deterministic force to a system 
with the additive noise \cite{che}. From the physical point of view, the double-well potential 
is of particular interest since it allows us to study confined systems and, in particular, 
to model the barrier penetration. The double-well system was examined predominantly 
for the Gaussian case \cite{kram}; then the first passage 
time problem resolves itself to a dynamical equation with ordinary boundary conditions \cite{cox}. 
The barrier penetration for the L\'evy flights was considered in Ref. \cite{dit,dyb2}. 
New effects emerge when one introduces -- in addition to the double-well potential -- 
a time-dependent periodic force. A rate of the jumping between the wells depends nonlinearly on 
the noise intensity and then a tuning of that intensity yields a matching of two characteristic 
frequencies: the noise-induced rate and a frequency of the deterministic, periodic force. That 
phenomenon, called a stochastic resonance \cite{benz}, is well-known for the Gaussian noise 
\cite{gamm1} and was analysed also for the multiplicative case \cite{gamm}. If the 
additive noise involves the L\'evy flights the stochastic resonance also emerges; 
it was observed in systems with both the double- \cite{kos,dybr} and single-well 
potential \cite{dybr1}. 

In this paper, we address a problem of the resonant switching between 
potential wells for the case of the multiplicative L\'evy-stable 
stochastic stimulation. We demonstrate to what extend properties of the system 
are modified, in respect to the known cases, when one introduces long jumps and 
takes into account a dependence of the noise on position for both SI and II. 
In Sec.II the system, which includes the bistable potential and the oscillatory force, 
is defined and a mean first passage time calculated. Existence of the stochastic resonance 
is demonstrated, and dependence of the signal amplitude on the system parameters 
discussed, in Sec.III. 

\section{Probability distributions and mean first passage time}

A general stable L\'evy distribution is expressed in terms of three 
parameters: the stability index $\alpha$ ($0<\alpha\le2$), which determines the tail of the 
distribution, the asymmetry parameter (skewness) and the translation parameter. Since, 
in the following, we restrict our analysis to the symmetric distributions, 
the index $\alpha$ completely determines the driving noise. The characteristic 
function for this case reads 
\begin{equation}
\label{lev}
{\widetilde p_\xi}(k)=\exp(-\sigma^\alpha|k|^\alpha), 
\end{equation}
where $\sigma$ measures the apparent width of the distribution. If $\alpha=2$ 
Eq.(\ref{lev}) represents the normal distribution, otherwise the tails are algebraic, 
$\sim |\xi|^{-1-\alpha}$, and the stochastic trajectory is discontinuous in a sense 
of the Lindeberg condition \cite{gar}. The variance is infinite, as well as all 
higher moments. 

We consider a motion in the time-dependent bistable potential 
\begin{equation}
\label{pot}
V(x,t)=-\frac{a}{2}x^2+\frac{b}{4}x^4-A_0x\cos(\omega_0t)
\end{equation}
where the amplitude $A_0$ and the frequency $\omega_0$ are constant. 
The system is driven by the noise $G(x)\eta(t)$ and 
we assume that increments of the noise $\eta(t)$ are distributed 
according to the characteristic function (\ref{lev}). 
Then the Langevin equation is of the form
\begin{equation}
\label{la}
\dot x=-\partial V(x,t)/\partial x+G(x)\eta(t). 
\end{equation}
Intensity of the noise algebraically depends on the position, 
\begin{equation}
\label{godx}
G(x)=|x|^{-\theta},
\end{equation}
where $\theta>-1$. 
In this paper we take into account two interpretations of the multiplicative 
noise in Eq.(\ref{la}): the It\^o interpretation and the Stratonovich 
interpretation. Components of the discretized stochastic integral 
in II are defined by $G[x(t_{i-1})][\eta(t_i)-\eta(t_{i-1})]$, 
i.e. $G[x(t)]$ is calculated before the noise acts. In SI, in turn, 
one takes a middle point: $G[(x(t_{i-1})+x(t_i))/2][\eta(t_i)-\eta(t_{i-1})]$. 
A difference between both interpretations, as well as probability density distributions 
for some nonlinear potentials, is well-known for the Gaussian processes \cite{vkam,gra}. 
The Fokker-Planck equation, corresponding to Eq.(\ref{la}) in II, 
involves a fractional derivative and a variable diffusion coefficient \cite{sche}: 
\begin{equation}
\label{fpi}
\frac{\partial}{\partial t}p(x,t)=-\frac{\partial}{\partial x}
F(x)p(x,t)+\sigma^\alpha\frac{\partial^\alpha}{\partial |x|^\alpha}|G(x)|^\alpha p(x,t),  
\end{equation}
where $F(x)=-\partial V(x,t)/\partial x$. 

A technical advantage of SI consists in an admissibility of the variable change which 
allows us, in the one-dimensional case, to reduce Eq.(\ref{la}) 
to the Langevin equation with the additive noise. Validity of the ordinary rules 
of the calculus was rigorously proved for $\alpha=2$ \cite{gar}. 
It was confirmed for $\alpha<2$ by the numerical simulations but the linear case 
required introduction of a cut-off to the distribution \cite{sro1,sro2}. On the other hand, 
rules of the ordinary calculus apply to systems with the coloured noise and the white-noise 
limit exists which was demonstrated for the generalised Ornstein-Uhlenbeck process 
\cite{sron}. After the transformation 
\begin{equation}
\label{yodx}
y(x)=\frac{1}{1+\theta}|x|^{1+\theta}\hbox{sgn}x, 
\end{equation}
the Langevin equation takes the form 
\begin{equation}
\label{las}
\dot y=-\partial \hat V(y,t)/\partial y+\eta(t) 
\end{equation}
where the effective potential reads 
\begin{equation}
\label{potef}
\hat V(y,t)=-\frac{a}{2}(1+\theta)y^2+\frac{b}{2(2+\theta)}(1+\theta)^{2+2/(1+\theta)}y^{2+2/(1+\theta)}-
\frac{A_0}{1+2\theta}(1+\theta)^{1+\theta/(1+\theta)}|y|^{1+\theta/(1+\theta)}\cos(\omega_0 t).
\end{equation}
Wells of the time-independent part of the potential are positioned at 
$y_{m}=\pm (a/b)^{(1+\theta)/2}/(1+\theta)$ and their depth declines with $\theta$. 

Properties of the stochastic system are different for both interpretations of Eq.(\ref{la}). 
The case $V(x,t)=0$ was discussed in Ref.\cite{sro1}. For II, the distribution tail 
does not depend on $\theta$ -- it falls like $|x|^{-1-\alpha}$ -- whereas SI leads to 
the dependence $|x|^{-1-\alpha-\theta\alpha}$. As a consequence, variance can be finite 
for SI. Result for the potential $V(x,t)=-\lambda x^2/2$, where $\lambda>0$, is similar. 
\begin{center}
\begin{figure}
\includegraphics[width=11cm]{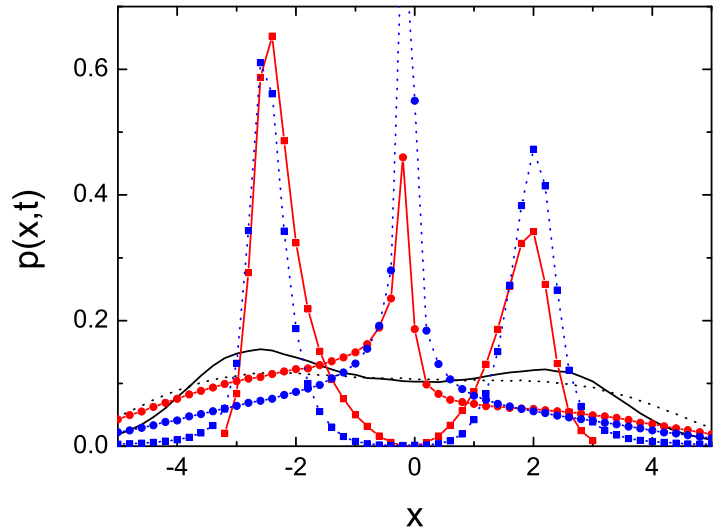}
\caption{(Colour online) The probability density distributions  
for II (blue dotted lines with symbols) and SI (red solid lines with symbols) 
calculated with $\alpha=1.5$ at $t=2$. 
The case $\theta=2$ is marked by squares and $\theta=-0.5$ by circles. 
Other parameters are $a=4$, $b=1$, $A_0=3$, $\omega_0=5$ and $\sigma=10$. 
The case $\theta=0$ with $\alpha=1.5$ and 2 are marked by the solid and dotted lines, 
respectively.}
\end{figure}
\end{center}

Presence of the nonlinear force modifies tails of the distribution, 
compared to linear case, making them steeper because, despite long jumps, 
the potential well prevents the particle 
from escaping to large distances. As a consequence, the variance may become finite. 
It is always the case for a quartic oscillator both with the additive noise \cite{che} and
the multiplicative noise in SI if $\theta>-1$ \cite{sro2}. 
Motion in the double-well potential with the $\alpha-$stable additive driving was considered 
by Ditlevsen \cite{dit}. The waiting and barrier penetration time -- defined in terms of the 
reflection and/or absorbing barriers -- is of particular interest for that system. 
An influence of the algebraic tails of the distribution for $\alpha<2$ on the average 
transition time between the wells becomes visible if width of the barrier 
is sufficiently large; then the transition may be caused by a single jump \cite{dit}. 
Otherwise the motion is dominated by a continuous superposition of many small 
jumps, similarly to the Gaussian case, and the transition time depends exponentially 
on the potential depth according to the Arrhenius formula. The mean first passage time 
displays a bell-shape as a function of $\alpha$: it rises with $\alpha$ up to 
$\alpha=1$ and then falls \cite{dyb2}. In the case of the multiplicative noise, 
a behaviour of the transition time as a function of system parameters depends on 
the specific interpretation of the stochastic integral \cite{sro2}. 
\begin{center}
\begin{figure}
\includegraphics[width=11cm]{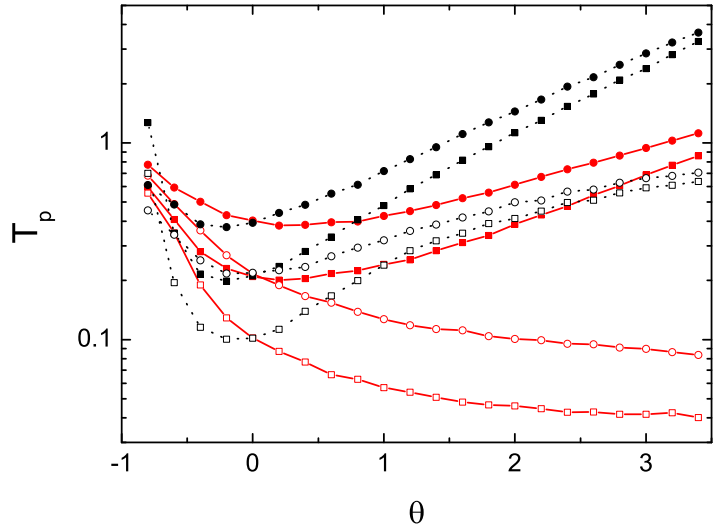}
\caption{(Colour online) Mean passage time from the left to the right well of the potential 
for II (black dotted lines) and SI (red solid lines) as a function of $\theta$. 
The case $\alpha=2$ is marked by squares 
and $\alpha=1.5$ by circles. Other parameters are $a=4$ (full symbols), $a=1$ 
(empty symbols), $b=1$, $A_0=3$, $\omega_0=5$ and $\sigma=10$.}
\end{figure}
\end{center}

A further generalisation of the above system includes the periodic driving force. 
For this system, Eq.(\ref{la}), we define the mean first passage time $T_p$ as 
the average time needed to pass for the first time from the right minimum 
of the potential to the left one. Therefore, we introduce an absorbing barrier 
at $x=x_{m}=-\sqrt{a/b}$ in a sense that trajectories for which $x(t)\le x_{m}$ 
are absorbed. The fractional Fokker-Planck equation corresponding to Eq.(\ref{la}) 
involves a nonlocal boundary condition, since the jumping particle may skip the boundary 
without hitting it, and the problem cannot be solved analytically \cite{dyb2}. 
In the present paper, the stochastic trajectories were calculated 
by the numerical integration of Eq.(\ref{la}). In the case of II, a simple Newtonian 
method was applied. For SI, in turn, we first transformed the variable $x\to y$, 
according to Eq.(\ref{yodx}), and then applied the Heun method to the equation with 
the additive noise, Eq(\ref{las}): 
\begin{eqnarray}
\label{heun}
{\tilde y}_{i+1}&=&y_i+h\hat F(y_i,t_i)\\
y_{i+1}&=&y_i+\frac{h}{2}[\hat F(y_i,t_i)+\hat F({\tilde y}_{i+1},t_{i+1})]+h^{1/\alpha}\xi_i\nonumber
\end{eqnarray}
where $\hat F(y,t)=-\partial \hat V(y,t)/\partial y$. 
\begin{center}
\begin{figure}
\includegraphics[width=11cm]{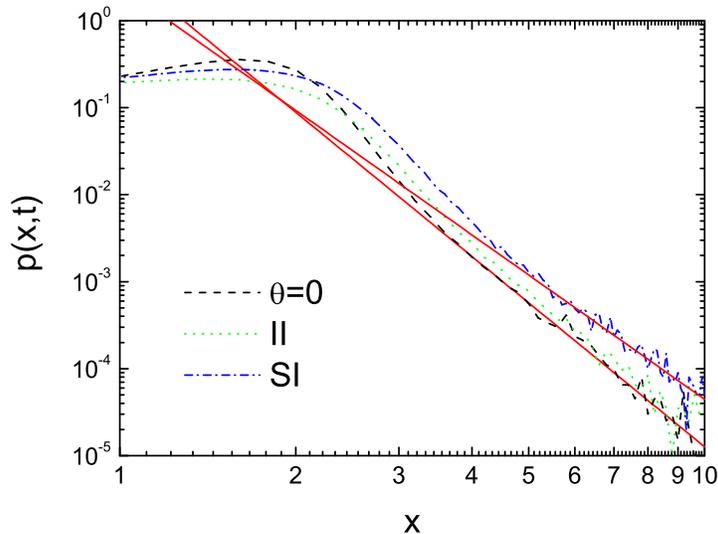}
\caption{(Colour online) The probability density distributions for both interpretations 
calculated at $t=2$ with the parameters: $\alpha=1.5$, $a=4$, $b=1$, $A_0=3$, 
$\omega_0=5$ and $\sigma=2$. Cases with the multiplicative noise were calculated with 
$\theta=-0.5$. The red solid lines are of the form $x^{-\mu}$ with 
$\mu=5.5$ and 4.75.}
\end{figure}
\end{center}

The multiplicative term $G(x)$ qualitatively modifies the distribution. First, we consider 
the case without the absorbing barrier. The distribution possesses 
two peaks, at negative and positive $x$, corresponding to the well positions. Their 
height decreases with $\sigma$ and, for $\theta<0$, they vanish altogether. Instead, 
a peak at the origin emerges as a result of increased contribution from the interwell 
motion. Fig.1 presents the probability density distributions 
obtained by a numerical integration of Eq.(\ref{la}) with the initial condition 
$p(x,0)=\delta(x-x_m)$. A relatively large noise intensity has been applied 
and the interwell motion prevails. In contrast to the case $\theta=0$, 
the distribution is strongly anisotropic for both interpretations.  
Since the noise vanishes at the origin for $\theta<0$, the particle abides near 
the top of the barrier. On the other hand, for the case $\theta>0$ both peaks are present 
even for a large $\sigma$. Those conclusions apply for both II and SI. 
The case $\theta=0$ (additive noise) exhibits small maxima at the well positions for 
$\alpha=1.5$ whereas the Gaussian case is almost uniform in a wide range of the argument. 
\begin{center}
\begin{figure}
\includegraphics[width=19cm]{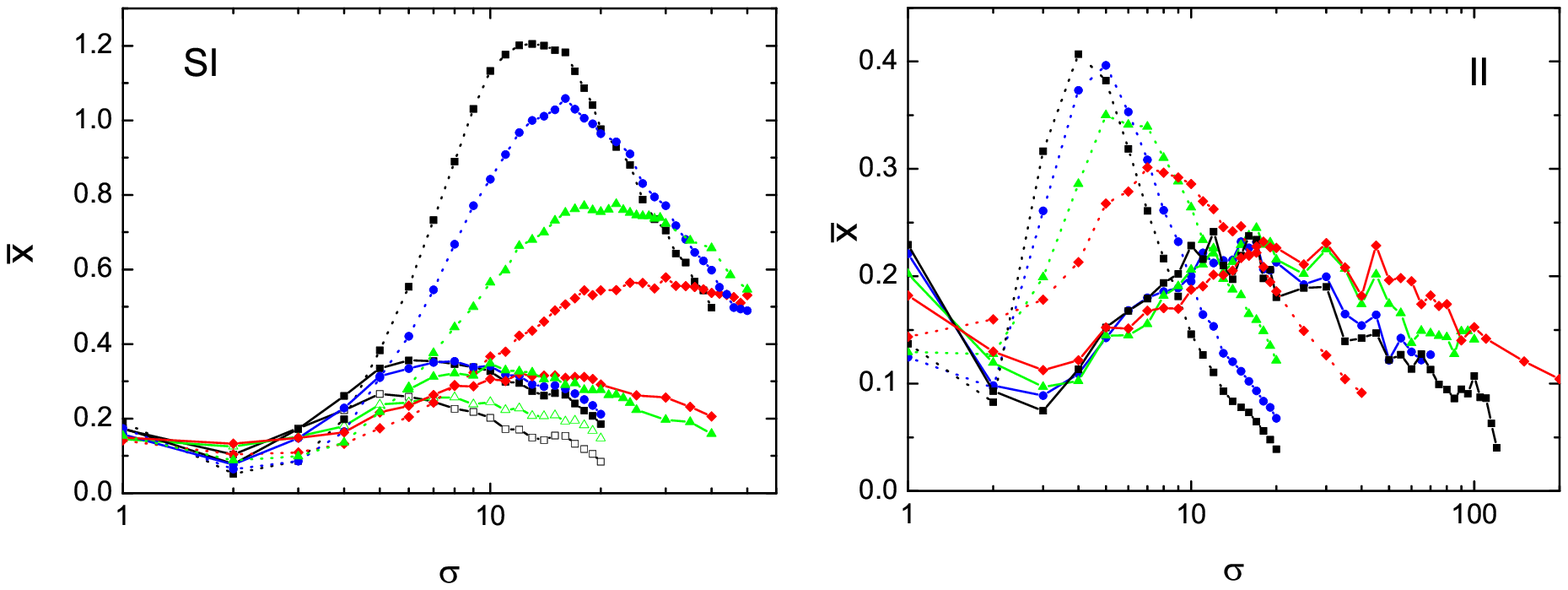}
\caption{(Colour online) $\bar x$ as a function of $\sigma$ for $\theta=1.1$ (solid lines) 
and $-0.6$ (dotted lines). The following values of $\alpha$ are presented: 2 (black squares),
1.8 (blue circles), 1.5 (green triangles) and 1.2 (red diamonds). Lines with empty symbols 
corresponds to $\theta=0$. Other parameters are $a=4$, $b=1$, $A_0=3$, $\omega_0=5$. Averaging 
was performed over $2\times10^4$ events.}
\end{figure}
\end{center}

The mean passage time as a function of $\theta$ for various configurations 
is presented in Fig.2. Results depend both on the stability index $\alpha$ and 
on the particular interpretation of the stochastic integral. Moreover, a sensitivity 
on the potential geometry is visible. First, let us consider the case 
of the relatively wide barrier ($a=4$). Depth of the effective potential for SI, 
Eq.(\ref{potef}), diminishes with $\theta$ and one could expect a smaller $T_p$. 
However, we know from Ref.\cite{dit} that, for a system with the additive 
noise, $T_p$ depends rather on the width of the barrier than on the potential depth 
if jumps determine the barrier penetration. We observe a similar effect: position 
of the effective well reaches its minimum at $\theta_m=2/\ln(a/b)-1$ which value 
corresponds to the minimum of $T_p$ in the figure. Nevertheless, there is no 
qualitative difference between the cases $\alpha=1.5$ and 2. Shape of the curves 
$T_p(\theta)$ for II is similar but $T_p$ rises faster for $\theta>0$, as a result 
of a strong noise damping. The passage time as a function of $\theta$ for $a=1$ 
is different than for $a=4$ because then $y_m(\theta)=1/(1+\theta)$, $\theta_m$ 
does not exist and $T_p(\theta)$ monotonically falls. Since $\theta>-1$, 
such a behaviour is expected for all narrow barriers, namely if $a\le b$. 
Obviously, $T_p$ in Fig.2 decreases with $\alpha$ because the noise intensity 
$\sim\sigma^\alpha$. 

\section{Stochastic resonance}

The rate of jumping between potential wells due to the noise can match 
the frequency $\omega_0$ of the deterministic oscillatory force 
revealing a resonant pattern. This phenomenon has been extensively studied for 
the Gaussian processes and observed also for the general 
L\'evy statistics of the noise. It was demonstrated in Ref.\cite{kos} that 
the signal-to-noise ratio exhibits a resonant amplification for many 
systems characterised by the long jumps. The spectral 
amplification and dynamic hysteresis loops were calculated for the asymmetric L\'evy 
noise with various $\alpha$ and skewness parameter \cite{dybr}; 
the amplification dwindles with decreasing $\alpha$. 
On the other hand, the stochastic resonance phenomenon for a linear multiplicative 
Gaussian noise was studied in Ref.\cite{gamm} and attributed to the transition from 
an intrawell to an interwell modulated dynamics. 
\begin{center}
\begin{figure}
\includegraphics[width=19cm]{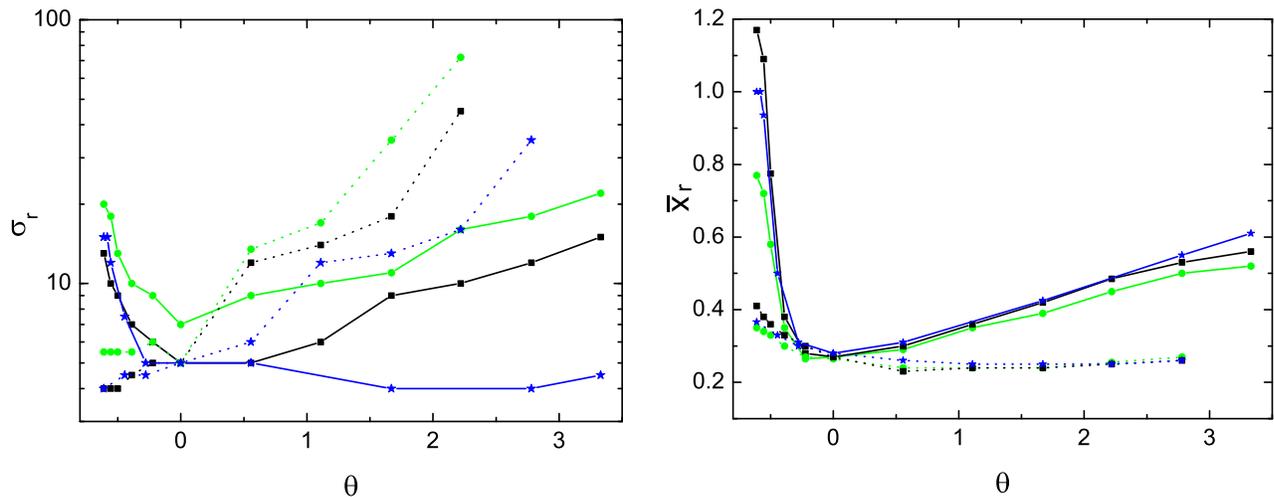}
\caption{(Colour online) Left part: $\sigma_r$ versus $\theta$ for SI (solid lines) and II (dotted lines). 
The case $a=4$ was calculated with $\alpha=2$ (black squares) and 1.5 (green circles), 
the case $a=1$ with $\alpha=1.8$ (blue stars). Other parameters are the same as in Fig.4. 
Right part: height of the peak versus $\theta$.}
\end{figure}
\end{center}

A useful quantity to describe a stochastic resonance is a correlation between 
the output process $x(t)$ and the periodic driving, $\bar x$. It equals the first Fourier 
coefficient which follows from the expression \cite{jun}
\begin{equation}
\label{wsp}
M_n=\frac{\omega_0}{\pi}\int_{-\infty}^\infty dx\int_0^T xp(x,t)\exp(in\omega_0t)dt,
\end{equation} 
where $T=2\pi/\omega_0$ and $p(x,t)$ results from Eq.(\ref{la}). The density $p(x,t)$ 
is asymptotically periodic and an averaging over the period produces a stationary distribution. 
The spectral amplification at the frequency $\omega_0$ is given by a ratio of the integrated 
power stored in spikes of the power spectrum to the total power carried by 
the oscillatory force, $\eta=4(\bar x/A_0)^2$, 
where $\bar x=\hbox{Re} M_1$ \cite{gamm1}. The signal-to-noise ratio (SNR) also can 
be expressed by $M_1$: SNR$=4\pi|M_1|^2/S_N^0(\omega_0)$ where $S_N^0$ is 
the total power spectral density. The latter quantity is directly related to the autocorrelation 
function and its existence requires the finite variance. That condition is satisfied for systems 
with the quartic potential in SI since then the tail $\sim |x|^{-3-\alpha-\alpha\theta}$ \cite{sro2}. 
On the other hand, an analysis for both the free particle and the linear force indicates that in II 
shape of the distribution tails does not depend on $\theta$. Fig.3 shows that 
also for the general system (\ref{la}) the tails fall faster than for the simple L\'evy motion: 
the slope rises with $\theta$ for SI whereas it does not depend on $\theta$ for II. 
Therefore variance remains finite when L\'evy flights are included and the spectral quantities, 
$\eta$ and SNR, are finite and well determined. 

For small amplitudes and large time, a response of the system to the periodic stimulation 
is proportional to $\bar x$, $\langle x(t)\rangle_{as}=\bar x\cos(\omega_0t-\bar\phi)$, 
where $\bar\phi$ is a phase lag \cite{gamm1}. The stochastic resonance is usually detected 
by plotting $\bar x$ as a function of the stochastic stimulation $\sigma$. That function 
does not monotonically diminish but exhibits a bell-shaped structure maximum of which 
corresponds to a matching condition of two frequencies: $\omega_0$ and a rate of 
the stochastic jumping, $1/T_p$. Then we can expect that a position of the resonance 
is directly related to the mean passage time and satisfies the approximate condition 
\begin{equation}
\label{warres}
\omega_0T_p(\sigma,\alpha,\theta)\sim1. 
\end{equation}
Hight of the peak rises with the amplitude $A_0$ and diminishes with the frequency $\omega_0$. 
In the following, we discuss an emergence of the stochastic resonance in the system (\ref{la}) 
and demonstrate how its position and intensity depends 
on $\alpha$, $\theta$, as well as geometry of the potential. 
\begin{center}
\begin{figure}
\includegraphics[width=19cm]{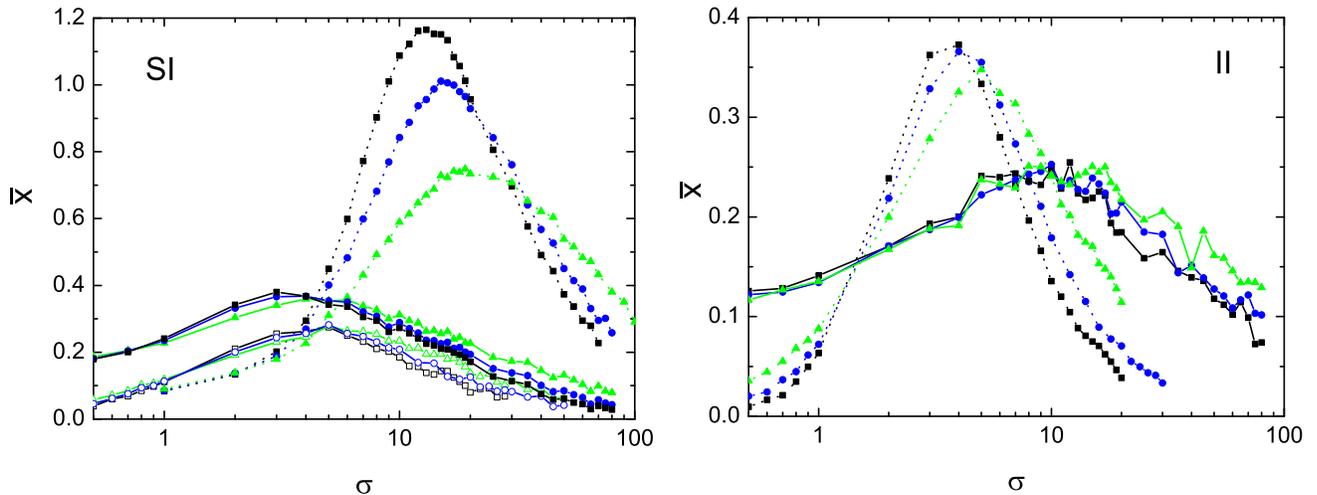}
\caption{(Colour online) The same as Fig.4 but for $a=1$ and $\alpha=2$, 1.8 and 1.5.}
\end{figure}
\end{center}

Fig.4 presents results of the calculations: a dependence of $\bar x$ on $\sigma$ 
for two values of $\theta$, both positive and negative; comparison with the additive 
noise is included. The resonance structure is visible for all the presented cases. 
Remarkable differences emerge when one compares both interpretations. 
For SI, $\bar x$ is large if $\theta<0$ and a position of the peak, $\sigma_r$, 
occurs at larger noise intensity than for $\theta>0$ since then the effective barrier is wider. 
The resonance for the positive $\theta$ emerges at roughly the same $\sigma$ as for $\theta=0$. 
The peak is much lower for II if $\theta<0$ and corresponds to a smaller $\sigma$ than 
for $\theta>0$. The latter case, governed by Eq.(\ref{fpi}), is characterised by a weak noise 
intensity when the particle remains far from the origin and that effect must be compensated 
by a larger $\sigma$. Dependence on $\alpha$ is similar for both interpretations: 
$\sigma_r$ shifts towards smaller $\sigma$ and $\bar x_r=\bar x(\sigma_r)$ rises with $\alpha$
for $\theta<0$. $\sigma_r$ falls with $\alpha$ also for $\theta>0$ but this effect is much 
weaker, especially for II. A dependence of $\sigma_r$ on $\theta$ 
is presented in Fig.5 for $\alpha=1.5$ and 2. The difference between both interpretations 
is especially pronounced for $\theta<0$ when $\sigma_r$ remains constant for II, 
in contrast to SI. For $\theta>0$, all presented cases exhibit a monotonically 
increasing function $\sigma_r(\theta)$. Height of the resonance, $\bar x_r$, is also presented 
in Fig.5 as a function of $\theta$. This dependence is very weak for II, in contrast to SI for which  
we observe a pattern similar to $\sigma_r(\theta)$ with a minimum at the same point. 
Results for both values of $\alpha$ are almost the same. An apparent similarity between 
Fig.2 and 5, which is especially striking for SI, is a consequence of the condition (\ref{warres}); 
since the curves $T_p(\theta)$ for different $\sigma$ are, approximately, parallel to each 
other, they mimic the dependence $\sigma_r(\theta)$. 

A change of the potential shape qualitatively modifies $T_p$ and one can expect 
differences also for the stochastic resonance. This conclusion is illustrated 
in Fig.5 which compares the case of the wide barrier ($a=4$) with the case 
of the narrow barrier ($a=1$) for both interpretations. Though results for II 
do not differ substantially -- $\sigma_r$ rises for both geometries -- the curve 
for SI and $a=1$ is flat resembling a similar behaviour of $T_p$, cf. Fig.2. 
Comparison of the peak heights is also presented in Fig.5. We conclude that 
the amplitude is not sensitive not only to $\alpha$ but also to the barrier width, 
in contrast to peaks' position. The amplification is much stronger for SI than 
for II and the dependence $\bar x_r(\theta)$ is determined mainly by a specific 
interpretation of the stochastic integral. Curves $\bar x(\sigma)$ for the narrow 
barrier, shown in Fig.6, are similar to the case $a=4$ (Fig.4); difference 
resolves itself to a smaller $\sigma_r$ for SI if $\theta\ge0$.

\section{Conclusions}

Properties of systems containing stochastic force, which is characterised 
by the stable distribution with the heavy tails, differ from those for the normal 
distribution possessing, in particular, the infinite variance. Systems with 
the multiplicative noise exhibit a more complicated behaviour and the variance 
may be finite, despite the algebraic tails. Nonlinear potentials, in turn, makes 
the distribution tails steeper and may invoke an effective trapping. 
In such systems motion is limited in space and asymptotics of the driving noise 
plays a relatively minor role, compared to the diffusion problems. 
Motion in the double-well potential is a simple example. 

We have discussed such a system: the anharmonic oscillator with the periodic 
modulation and the stochastic stimulation in a form of the stable, multiplicative 
process. The multiplicative factor, assumed in the algebraic form, makes 
the distribution anisotropic; the point $x=0$ becomes an attractor for $\theta<0$
since noise vanishes at the origin. The mean first passage time depends on a particular 
interpretation of the stochastic integral; for SI it is determined rather 
by the size of the effective potential than by the effective well depth. That conclusion, 
however, is valid only for a relatively wide barrier and then the geometry is 
important for the dependence $T_p(\theta)$. The passage time smoothly rises 
with $\alpha$ for both geometries and no qualitative difference in respect to 
$\alpha=2$ has been observed. 

The system (\ref{la}) possesses two characteristic frequencies, $\omega_0$ and $1/T_p$, 
and, when they match, the resonance occurs. The resonance shape of the dependence 
$\bar x(\sigma)$ has been observed for all analysed cases. Since the variance is finite 
also for processes with the L\'evy flights, both the spectral amplification $\eta$ and SNR 
are well determined. For SI, height of the resonance is sensitive on $\theta$ and the 
amplification rapidly rises when $\theta$ becomes negative. 
Position of the resonance, in turn, mimics the dependence $T_p(\theta)$. 
Therefore, size of the barrier is important for the resonance position: $\sigma_r$ 
is almost constant for SI if $a=1$, in contrast to the case $a=4$. 
We finally conclude that -- though height and position of the peaks depend 
on $\alpha$ -- the bistable system with the L\'evy flights has similar properties 
as the case of the Gaussian processes. On the other hand, dependence of the noise 
on the process value essentially modifies -- differently for each interpretation 
of the stochastic integral -- the distributions, the passage time and properties 
of the stochastic resonance.

\end{document}